\documentclass[pra, twocolumn, floatfix]{revtex4-2}
\usepackage{graphicx}
\usepackage{amsmath, amsfonts, amssymb, bm,color}

\def\ifa{Institute of Applied Physics, Moldova State University,
Academiei str. 5, MD-2028 Chi\c{s}in\u{a}u, Moldova}

\begin{document}
\title{Dipole-dipole interacting two-level emitters in a moderately intense laser field}

\author{Profirie Bardetski }
\affiliation{\ifa}

\author{Mihai A. Macovei }
\email{mihai.macovei@ifa.usm.md}
\affiliation{\ifa}

\date{\today}
\begin{abstract}
We investigate the resonance fluorescence features of a small ensemble of closely packed and moderately laser pumped 
two-level emitters at resonance. The mean distance between any two-level radiators is smaller than the corresponding 
emission wavelength, such that the dipole-dipole interactions are not negligible. We have found that under the secular 
approximation, the collective resonance fluorescence spectrum consists of $2N+1$ spectral lines, where $N$ is the 
number of emitters from the sample. The $2N$ spectral sidebands, symmetrically located around the generalized 
Rabi frequency with respect to the central line at the laser frequency, are distinguishable if the dipole-dipole coupling 
strength is larger than the collective spontaneous decay rate. This way, one can estimate the radiators number within 
the ensemble via measuring of the spontaneously scattered collective resonance fluorescence spectrum. Contrary, if 
the dipole-dipole coupling is of the order of or smaller than the cooperative spontaneous decay rate, but still non-negligible, 
the spectrum turns into a Mollow-like fluorescence spectrum, where the two sidebands spectral lines broadens, proportional to 
the dipole-dipole coupling strength, respectively.
\end{abstract}
\maketitle

\section{Introduction}
Resonance fluorescence of two-level atoms was extensively studied in the scientific literature and significant results were already 
obtained and reported \cite{walther,tan,ScZb}. An important issue of this research topic is the well-known Mollow spectrum 
\cite{mollsp} and, probably, there is no need to justify its relevance. It was observed in a wide range of systems, like e.g. 
atomic beams \cite{mollB}, single molecules \cite{mollM},  quantum dots \cite{mollQD} or wells \cite{exc}, or cold atoms 
\cite{mollCA}. The spectral lines features were investigated earlier as well \cite{raut} and, more recently, for higher frequency 
ranges \cite{octav} or for polar molecules \cite{pasp}, respectively.

When many two-level emitters are considered in the resonance fluorescence phenomena, things can change depending on 
the mean inter-particle separations \cite{dicke,lehm,gsag,supm1,thrF,supm2,gxLB,puri,ficek,reww,rew_all}. If the atomic ensemble 
is concentrated in a space volume with linear dimensions smaller than the photon emission wavelength, the collective resonance 
fluorescence spectrum may consists from multiple spectral lines with higher intensities and various spectral widths, depending 
on the atomic sizes and external coherent pumping strengths \cite{colSP1,helenf1,kilin2N,helen,hars,colSP2,fkT,CorR1,twat1,
CorR2,tqkl,twat2,mek,brNat,alej,elnaz}. 
In this regard, recent experiments include measurements of fluorescence emission spectra of few strongly driven atoms using 
an optical nanofiber \cite{expFM}, agreeing well with the Mollow spectrum. Observations of broadening of the spectral line, a 
small redshift and a strong suppression of the scattered light, with respect to the non-interacting atomic case, in driven 
dipole-dipole interacting atoms was reported as well, in Ref.~\cite{ddExpB}. Furthermore, in a dilute cloud of  strongly driven 
two-level emitters, the Mollow triplet is affected by cooperativity too and exhibits asymmetrical behaviours under experimental
conditions \cite{KaisMol}. While finding analytical expressions for emission or absorption spectra in a two-level many-atom ensemble 
is a challenging task, it was demonstrated in Ref.~\cite{FrdN} that these quantities may assume a simple mathematical structure 
if the atoms are arranged in a particular geometrical configuration.

Thus, motivated by recent advances in experimental research dealing with cooperative interactions among many two-level emitters, 
here we investigate the collective interaction of a small and closely packed ensemble of $N$ two-level motionless emitters with an 
externally applied coherent laser wave. The inter-particle separations are less than the corresponding emission wavelength, 
therefore, the dipole-dipole interactions among the two-level radiators are included and can play a relevant role under
specific conditions. Particularly, we focus on a situation when the Rabi frequency, arising due to the ensemble interaction 
with the resonantly applied coherent laser field, is larger than the collective spontaneous decay rate, respectively. On the 
other hand, it is being commensurable to the dipole-dipole interaction strength, but still bigger. Under these conditions, 
we have analytically calculated the collective resonance fluorescence spectrum, spontaneously scattered by the laser-driven 
dipole-dipole interacting two-level radiators, and have found that it consists of $2N+1$ spectral lines. Each of $N$ spectral 
sidebands are symmetrically generated, with respect to the central spectral line at the laser frequency, around the Rabi 
frequency, respectively. Furthermore, these spectral sidebands become distinguishable if the dipole-dipole coupling strength,
assumed constant for all involved atomic pairs, is larger than the collective spontaneous decay. Actually, the sidebands occurs 
due to transitions among the $N+1$ symmetrical collective Dicke states formed from individual laser-atom dressed states. In 
the opposite case, i.e. when the dipole-dipole coupling is similar to or less than the cooperative spontaneous decay rate, 
the fluorescence spectrum turns into a three-line Mollow-like spectrum, where the spectral widths of the sidebands broadens, 
proportional to the dipole-dipole interaction coupling strength. As a possible application of the reported results, one can 
estimate the emitters number by measuring the incoherent collective resonance fluorescence spectrum. Alternatively, one 
can extract the dipole-dipole coupling strength as well as the mean-distance among the closely spaced two-level radiators 
in a small laser-pumped ensemble, because the frequency interval among the $N$ spectral sidebands is given by the scaled 
dipole-dipole coupling, which in turn is inversely proportional to the cubic mean inter-particle separations, respectively. 
Notice a recent experiment \cite{pDM}, where the laser-driven two-level Dicke model was realized experimentally. There, 
the earlier theoretically predicted non-equilibrium superradiant phase transition in free space, based on a laser pumped 
two-level Dicke-like ensemble where all the inter-particle couplings are considered identical while the emitters motionless, 
was successfully demonstrated. This makes our findings experimentally achievable in principle.
\begin{figure}[t]
\includegraphics[height =3.3cm]{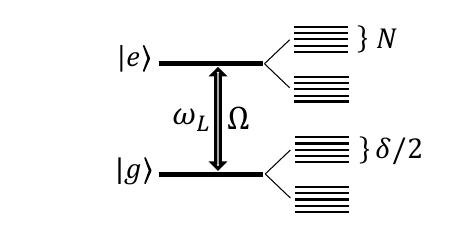}
\includegraphics[height =3.3cm]{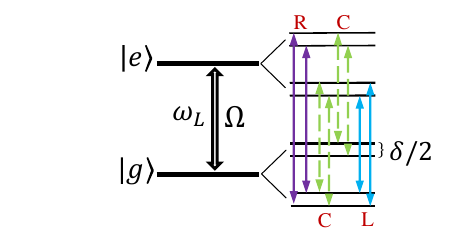}
\begin{picture}(0,0)
\put(-190,160){(a)}
\put(-190,70){(b)}
\end{picture}
\caption{\label{fig-0ab}
The schematic of the involved laser-atoms dressed-state model. (a) An external coherent field of 
frequency $\omega_{L}$ resonantly drives $N$ two-level radiators on transition $|e\rangle \leftrightarrow |g\rangle$, 
with $\Omega$ being the corresponding Rabi frequency. The bare atomic levels are dressed by the applied laser field 
leading to traditional laser-atom dressed-states splitting proportional to $2\Omega$, i.e. the dynamical Stark effect. 
These dressed energy levels would additionally split due to the dipole-dipole interaction, $\delta$, among the two-level 
emitters in $N$ sublevels, respectively. (b) The cooperative atom-laser dressed-states transitions responsible for the 
resonance fluorescence spectrum of a two-atom Dicke-like sample, $N=2$. Here $R$ shows the transitions leading to 
right spectral bands at $\omega_{L}+2\Omega \pm \delta/2$, whereas $L$ those corresponding to the transitions for 
the left spectral lines at $\omega_{L}-2\Omega \pm \delta/2$. Respectively, $C$ depicts the dressed-state transitions 
for the central line at $\omega_{L}$, see Fig.~(\ref{fig-1}) and Appendix \ref{appB}.}
\end{figure}
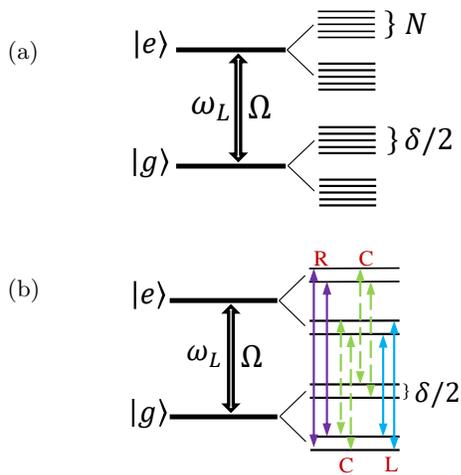

This paper is organized as follows. In Sec.~\ref{theo} we describe the system of interest and the analytical approach based on the 
master equation formalism, while in Sec.~\ref{SP} we calculate the collective resonance fluorescence spectrum spontaneously 
scattered by the laser-pumped two-level emitters. Sec.~\ref{RD} presents and analyses the obtained results. The article concludes 
with a summary given in Sec.~\ref{sum}.

\section{Theoretical framework \label{theo}}
The system of interest consists from an ensemble of externally coherently laser pumped and dipole-dipole interacting 
$N$ two-level emitters, within the Dicke limit \cite{dicke}. The Hamiltonian describing this system in the dipole and 
rotating-wave approximations \cite{gsag,supm1,thrF,supm2,gxLB,puri,ficek,reww}, in a frame rotating at the laser frequency 
$\omega_{L}$, is as follows:
\begin{eqnarray}
H &=&\sum_{k}\hbar(\omega_{k}-\omega_{L})a^{\dagger}_{k}a_{k} + \hbar\Delta S_{z} + \hbar\Omega(S^{+}+S^{-})
\nonumber \\
&-&\hbar\tilde\delta S^{+}S^{-} + i\sum_{k}(\vec g_{k}\cdot \vec d)(a^{\dagger}_{k}S^{-}-S^{+}a_{k}), \label{dHm}
\end{eqnarray}
where $\Delta= \omega_{0} -\omega_{L} + \tilde \delta$, with $\omega_{0}$ being the emitters transition frequency. Here,
the collective atomic operators $S^{+} = \sum^{N}_{j=1}|e\rangle_{j}{}_{j}\langle g|$ and $S^{-}=[S^{+}]^{\dagger}$ 
obey the usual commutation relations for su(2) algebra, namely, $[S^{+},S^{-}] =2S_{z}$ and $[S_{z},S^{\pm}]=\pm S^{\pm}$, 
where $S_{z} = \sum^{N}_{j=1}(|e\rangle_{j}{}_{j}\langle e| - |g\rangle_{j}{}_{j}\langle g|)/2$ is the bare-state inversion operator. 
$|e\rangle_{j}$ and $|g\rangle_{j}$ are the excited and ground state of the emitter $j$, respectively, while $a^{\dagger}_{k}$ 
and $a_{k}$ are the creation and the annihilation operators of the environmental electromagnetic field (EMF) vacuum reservoir which 
satisfy the standard bosonic commutation relations, i.e., $[a_{k}, a^{\dagger}_{k'}] = \delta_{kk'}$, and $[a_{k},a_{k'}]$ = 
$[a^{\dagger}_{k},a^{\dagger}_{k'}] = 0$ \cite{gsag,supm1,thrF,supm2,gxLB,puri,ficek,reww}.  In the Hamiltonian (\ref{dHm}), the 
free energies of the EMF vacuum modes and atomic subsystems are given by the first two terms of the Hamiltonian. The third and 
fifth components account for the laser as well as the EMF surrounding vacuum modes interactions with the two-level emitters, 
respectively. There $\Omega$ is the corresponding Rabi frequency due to the external applied coherent laser field, whereas 
$\vec g_{k}$=$\sqrt{2\pi\hbar\omega_{k}/V}\vec e_{p}$ is the coupling strength among the few-level atoms and the EMF vacuum 
modes. Here $\vec e_{p}$ is the photon polarization vector with $p \in \{1,2\}$ and $V$ is the quantization volume. 
The fourth term of the Hamiltonian (\ref{dHm}) describes the dipole-dipole interactions among the two-level emitters, 
obtained from the dipole-dipole Hamiltonian $H_{dd}=\hbar\delta\sum_{j\not=l}S^{+}_{j}S^{-}_{l}$ \cite{gsag}. Here $\delta$ is the 
dipole-dipole coupling strength taken equal for any atomic pair. This is a reasonable approximation in the Dicke limit via assuming a densely 
packed atomic ensemble with its linear dimensions much smaller than the photon emission wavelength $\lambda$. In this case, the 
dipole-dipole coupling is mainly being proportional to $\delta \sim d^{2}/r^{3}$ \cite{gxLB,puri,ficek,reww}, where $r$ is the mean 
distance among any atomic pair characterized by dipole $d$. Alternatively, one can assume a Gaussian-distributed atomic cloud to obtain 
an averaged dipole-dipole coupling strength $\delta$, see e.g. \cite{KaisMol}. Observing that $S^{+}S^{-}$
=$\sum_{j\not=l}S^{+}_{j}S^{-}_{l}$ + $\sum_{j}S_{zj}+N/2$, one obtains the following expression for the dipole-dipole interaction 
Hamiltonian: $H_{dd}=\hbar\tilde\delta(S^{+}S^{-} - S_{z})$, where the constant $N/2$ is being dropped. These components can 
be recognized in the Hamiltonian (\ref{dHm}). Notice that the Hamiltonian of many atoms is an additive function, i.e. it consists 
from a sum of individual Hamiltonians, describing separately each two-level emitter. From this reason, the dipole-dipole coupling 
strength $\delta$ was formally divided on $N-1$, i.e. $\tilde \delta$=$\delta/(N-1)$, because there are $\sum_{j\not=l} \to N(N-1)$ 
terms describing the dipole-dipole interacting atoms \cite{supm1}. 

Under the action of the laser field, the system is conveniently described using the dressed-state formalism \cite{tan,ScZb}:
$|e\rangle_{j}=\cos{\theta}|\tilde e\rangle_{j}-\sin{\theta}|\tilde g\rangle_{j}$ and $|g\rangle_{j}=\sin{\theta}
|\tilde e\rangle_{j}+\cos{\theta}|\tilde g\rangle_{j}$ with $\cot{2\theta}=\Delta/2\Omega$. The system 
Hamiltonian (\ref{dHm}) can be written then as $H=H_{0} + H_{I}$, where
\begin{eqnarray}
H_{0}&=& \sum_{k}\hbar(\omega_{k}-\omega_{L})a^{\dagger}_{k}a_{k} + \hbar\bar G R_{z} -
\hbar\bar\delta R^{+}R^{-}, \nonumber \\
H_{I}&=&i\sum_{k}(\vec g_{k}\cdot \vec d)\bigl\{a^{\dagger}_{k}(\sin{2\theta}R_{z}/2 + \cos^{2}{\theta}R^{-}
\nonumber \\
&-& \sin^{2}{\theta}R^{+})- H.c.\bigr\}, \label{drH}
\end{eqnarray}
with $\bar G= G + \tilde \delta(\sin^{4}{\theta}-\sin^{2}{2\theta}/2)$, $G=\sqrt{\Omega^{2}+(\Delta/2)^{2}}$ and $\bar \delta = 
\tilde \delta(\cos^{4}{\theta} + \sin^{4}{\theta}-\sin^{2}{2\theta})$. Here, in the dipole-dipole part of the Hamiltonian, we have 
neglected fastly oscillating terms proportional to $e^{\pm ikGt}$, $\{k = 2,4\}$, while supposing that $\Omega \gg \tilde \delta$, 
and have used the relation $R^{2}_{z}/4 + (R^{+}R^{-} + R^{-}R^{+})/2=j(j+1)$, where $j=N/2$. 
The new quasispin operators $R^{+}=\sum^{N}_{j=1}|\tilde e\rangle_{j}{}_{j}\langle \tilde g|$, $R^{-}=[R^{+}]^{\dagger}$ and 
$R_{z}=\sum^{N}_{j=1}(|\tilde e\rangle_{j}{}_{j}\langle \tilde e|-|\tilde g\rangle_{j}{}_{j}\langle \tilde g|)$ operate in the dressed-state 
picture and obey the commutation relations: $[R^{+},R^{-}]=R_{z}$ and $[R_{z},R^{\pm}]=\pm 2R^{\pm}$. In the interaction picture, 
given by the unitary transformation $U(t)=\exp{(iH_{0}t/\hbar)}$, one arrives at the interaction Hamiltonian, $H_{i}(t)=U(t)H_{I}U^{-1}(t)$, 
that is
\begin{eqnarray}
H_{i}(t)=i\sum_{k}(\vec g_{k}\cdot \vec d)a^{\dagger}_{k}\bar R^{-}(t)e^{i(\omega_{k}-\omega_{L})t} + H.c.,
\label{Hmi}
\end{eqnarray}
where
\begin{eqnarray}
\bar R^{-}(t)=\sin{2\theta}R_{z}/2+\cos^{2}{\theta}e^{-i\hat\omega t}R^{-}-\sin^{2}{\theta}
R^{+}e^{i\hat\omega t}, \nonumber
\end{eqnarray}
with $\hat\omega=2\bar G + \bar\delta R_{z}$ and $\bar R^{+}=[\bar R^{-}]^{\dagger}$.

The general form of the dressed master equation, in the interaction picture describing the atomic subsystem alone, is given by
\begin{eqnarray}
\frac{d}{dt}\rho(t) &+& \frac{i}{\hbar}[H_{a},\rho(t)] =\nonumber \\
&-& \frac{1}{\hbar^{2}}{\rm Tr_{f}}\biggl\{\int^{t}_{0}dt'
\bigl[H_{i}(t),[H_{i}(t'),\rho(t')]\bigr]\biggr\}, \label{mq}
\end{eqnarray}
where $H_{a}=\hbar\bar G R_{z}-\hbar\bar\delta R^{+}R^{-}$, and the notation ${\rm Tr_{f}}\{ \cdots \}$ means the trace over the 
vacuum EMF degrees of freedom. Substituting the Hamiltonian (\ref{Hmi}) in the above equation (\ref{mq}), after cumbersome but not 
difficult calculations, one arrives at the final master equation describing the atomic subsystem only
\begin{eqnarray}
&{}&\frac{d}{dt}\rho(t)+ \frac{i}{\hbar}[H_{a},\rho(t)] =-\frac{\Gamma_{0}}{8}\sin^{2}{2\theta}
[R_{z},R_{z}\rho]\nonumber \\
&-&\frac{1}{2}\cos^{4}{\theta}[R^{+},\hat\Gamma^{(+)}R^{-}\rho] - \frac{1}{2}\sin^{4}{\theta}
[R^{-},R^{+}\hat\Gamma^{(-)}\rho] \nonumber \\
&+& H.c., \label{fMeq}
\end{eqnarray}
where $\Gamma_{0}=\gamma(\omega_{L})$ and $\hat \Gamma^{(\pm)}=\gamma(\omega_{L}\pm \hat\omega)$ are the 
spontaneous decay rates, i.e. $\gamma(\omega)=2d^{2}\omega^{3}/(3\hbar c^{3})$, at frequencies $\omega_{L}$ and 
$\omega_{L} \pm \hat\omega$, respectively. Note that we have performed the secular approximation when obtaining 
Eq.~(\ref{fMeq}), that is, we neglected rapidly oscillating terms proportional to $e^{\pm ik\bar Gt}$, $\{k \in 2,4\}$, 
meaning generally that $2\bar G \gg \{N\gamma(\omega_{L}),\bar\delta\}$. One can observe that 
$\hat \Gamma^{(\pm)} \approx \Gamma_{0} \equiv \gamma$, since the eigenvalues of the dressed-state 
inversion operator vary within $\pm N$ and we consider that $\omega_{L} \gg 2\bar G \pm \bar\delta N$.

At resonance, when $\Delta=0$ or $\omega_{L}=\omega_{0} + \tilde \delta$, one has that $\theta=\pi/4$ and the master equation (\ref{fMeq}) 
possesses a steady-state solution
\begin{eqnarray}
\rho_{s}=\frac{\hat I}{N+1}, \label{ssol}
\end{eqnarray}
where $\hat I$ is the unity operator \cite{puri,colSP1}. We shall use the steady state solution (\ref{ssol}) in the next section, 
when calculating the resonance fluorescence spectrum of dipole-dipole interacting two-level atoms in a moderately intense and 
coherent laser field.

\section{The collective resonance fluorescence spectrum \label{SP}}
In the far-field limit, $R=|\vec R| \gg \lambda$, one can express the entire steady-state fluorescence spectrum via the collective atomic 
operators as
\begin{eqnarray}
S(\nu)=\Phi(R){\rm Re}\biggl\{\int^{\infty}_{0}d\tau e^{i(\nu-\omega_{L})\tau}
\langle S^{+}S^{-}(\tau)\rangle_{s}\biggr\},
\end{eqnarray}
where the subindex $s$ means steady-state. $\Phi(R)$ is a geometrical factor which we set
equal to unity in the following, while $\nu$ is the detected photon frequency. In the
dressed-state picture and at resonance, i.e. $\theta=\pi/4$, the fluorescence spectrum
transforms as follows in the secular approximation
\begin{eqnarray}
S(\nu)&=&{\rm Re}\int^{\infty}_{0}d\tau e^{i(\nu-\omega_{L})\tau}
\biggl\{\langle R_{z}R_{z}(\tau)\rangle_{s} + \langle R^{+}R^{-}(\tau)\rangle_{s} \nonumber \\
&+&\langle R^{-}R^{+}(\tau)\rangle_{s}\biggr\}/4. \label{sp1}
\end{eqnarray}
Now, using the master equation (\ref{fMeq}), one can obtain the time-dependences for the collective dressed-state atomic operators entering 
in the expression (\ref{sp1}) for the resonance fluorescence spectrum, namely,
\begin{eqnarray}
R_{z}(\tau) &=&  R_{z}e^{-\gamma\tau/2}, \nonumber \\
R^{-}(\tau) &=& e^{-(i\hat\omega + \hat\gamma_{c})\tau}R^{-}, \nonumber \\
R^{+}(\tau)&=& R^{+}e^{(i\hat\omega - \hat \gamma_{c})\tau}, \label{td}
\end{eqnarray}
where for $\theta=\pi/4$, $\hat\gamma_{c}=\gamma(1+2R^{-}R^{+})/4$, see Appendix \ref{appA}, and reminding that 
$\hat\omega=2\Omega - \tilde\delta(1+ R_{z})/2$. Inserting the time-solutions (\ref{td}) in (\ref{sp1}) and performing the 
integration and using the quantum regression theorem, one arrives at the following exact expression for the incoherent 
steady-state collective resonance fluorescence spectrum, that is,
\begin{eqnarray}
S(\nu) &=& \frac{1}{4}\biggl\{ \sum_{\pm}\sum^{N}_{n=0}I^{(\pm)}_{n}\frac{\gamma^{(\pm)}_{n}}{\gamma^{(\pm)2}_{n}
+ \bigl(\nu - \omega^{(\pm)}_{n}\bigr)^{2}} \nonumber \\
&+& I_{0}\frac{\gamma/2}{\bigl(\gamma/2\bigr)^{2} + \bigl(\nu - \omega_{L}\bigr)^{2}}\biggr\}. \label{fSp}
\end{eqnarray}
Here, the symbol $\sum_{\pm}$ indicates that both sets of spectral lines appearing with the plus sign and with the minus sign need 
to be incorporated in the sum. There, $I_{0}=N(N+2)/3$, $I^{(+)}_{n}=n(N-n+1)/\bigl(N+1\bigr)$, 
$I^{(-)}_{n}=(n+1)(N-n)/\bigl(N+1\bigr)$, $\gamma^{(+)}_{n}=\gamma\bigl(1+n(N-n+1)\bigr)/4$, $\gamma^{(-)}_{n}=\gamma\bigl(1+ 
(n+1)(N-n)\bigr)/4$ and $\omega^{(\pm)}_{n}$ = $\omega_{L} \pm 2\Omega \mp \tilde \delta(2n-N \mp 1)/2$, respectively. Note that 
the incoherent collective resonance fluorescence spectrum, given by the expression (\ref{fSp}), is valid under the secular approximation, i.e. 
$2\Omega > \delta \gg N\gamma$. Therefore, the corrections to the results obtained in the secular approximation are of the order of 
$(\delta/2\Omega)^{2}$.
\begin{figure}[t]
\includegraphics[width =7cm]{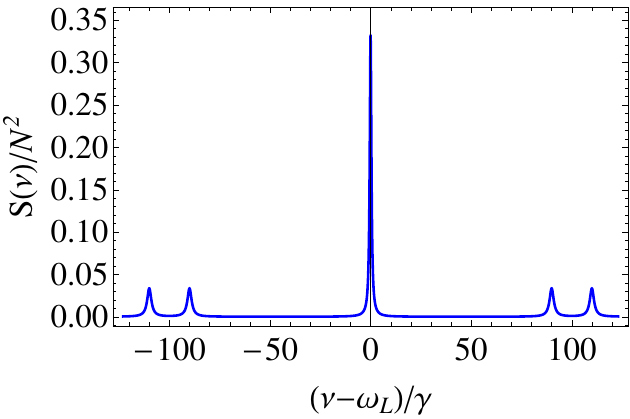}
\caption{\label{fig-1}
The scaled collective resonance fluorescence spectrum $S({\nu})/N^{2}$ as a function of $(\nu -\omega_{L})/\gamma$.
Here $N=2$, $2\Omega/\gamma=100$, $\Delta/\gamma=0$, and $\delta/(2\Omega)=0.2$, respectively.}
\end{figure}

In order to calculate the corresponding collective dressed-state correlators entering in the expression for resonance fluorescence spectrum, 
after inserting (\ref{td}) in (\ref{sp1}), we considered an atomic coherent state $|n\rangle$, which is a symmetrized $N$-atom state with 
$N – n$ particles in the lower dressed state $|\tilde g \rangle$ and $n$ atoms excited to the upper dressed state $|\tilde e\rangle$. We 
can calculate then the steady-state expectation values of any atomic correlators of interest, for $\theta=\pi/4$, using the steady-state 
solution (\ref{ssol}) of the master equation (\ref{fMeq}) as well as the relations: $R^{+}|n\rangle = \sqrt{(N-n)(n+1)}|n+1\rangle$, 
$R^{-}|n\rangle = \sqrt{n(N-n+1)}|n-1\rangle$ and $R_{z}|n\rangle = (2n-N)|n\rangle$. Particularly, if the dipole-dipole interactions 
are ignored, i.e. $\delta/2\Omega \to 0$, then one obtains the well-known expression for the incoherent collective resonance fluorescence 
spectrum. Actually, in this case, the corresponding time-dependent atomic operators can be obtained directly from the master equation (\ref{fMeq}), 
namely: $R_{z}(\tau)=R_{z}e^{-\gamma\tau/2}$, $R^{-}(\tau)=R^{-}e^{-(2i\Omega+3\gamma/4)\tau}$, and $R^{+}(\tau)=
R^{+}e^{(2i\Omega-3\gamma/4)\tau}$, to arrive at
\begin{eqnarray}
S(\nu) &=&\frac{1}{4}\biggl\{\sum_{\pm}I_{\pm}\frac{3\gamma/4}{\bigl(3\gamma/4\bigr)^{2} 
+ \bigl(\nu - \omega_{L} \mp 2\Omega\bigr)^{2}} \nonumber \\
&+& I_{0}\frac{\gamma/2}{\bigl(\gamma/2\bigr)^{2} + \bigl(\nu - \omega_{L}\bigr)^{2}}\biggr\}, \label{fSd0}
\end{eqnarray}
where $I_{\pm}=N(N+2)/6$, see e.g. \cite{colSP1,colSP2,tqkl}. Again here the symbol $\sum_{\pm}$ indicates that there are two spectral lines, 
one appearing with the plus sign and another one with the minus sign, respectively. The incoherent resonance fluorescence spectrum (\ref{fSd0}) 
turns into the famous single-atom Mollow spectrum \cite{mollsp}, if one sets $N=1$, with the central- and side-bands spectral widths equal to 
$\gamma/2$ and $3\gamma/4$, respectively. Moreover, in both cases, that is for the ratio $\delta/2\Omega \not =0$ but smaller than unity, 
or $\delta/2\Omega \to 0$, the collective resonance fluorescence spectra, i.e. the expressions (\ref{fSp}) and (\ref{fSd0}), are proportional 
to the squared number of involved two-level emitters, $S(\nu) \propto N^{2}$. Note that setting $N=2$ or $N=3$ in the general expression 
for the resonance fluorescence spectrum (\ref{fSp}), one gets identical expressions to those obtained by directly solving the master equation 
(\ref{fMeq}) for $N=2$ and $N=3$, which are given in the Appendix~\ref{appB} and the Appendix~\ref{appC}, i.e. Exp.~(\ref{toSp}) and 
Exp.~(\ref{thrSp}), respectively. Finally, the coherent part of the spectrum vanishes for the resonant laser pumping case, i.e. $\theta=\pi/4$, 
which is considered here, see also \cite{colSP2}. 
\begin{figure}[t]
\includegraphics[width =7cm]{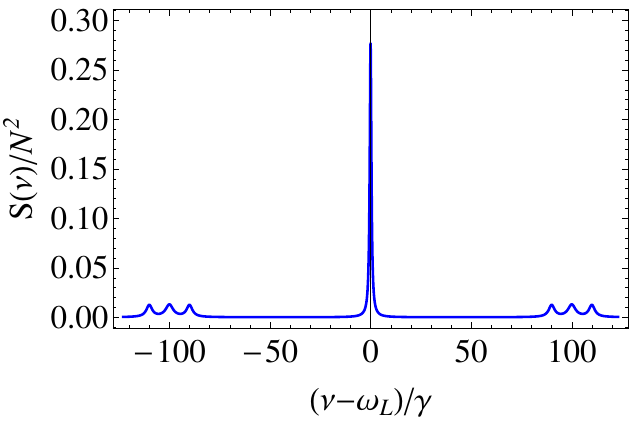}
\caption{\label{fig-2}
Same as in Fig.~(\ref{fig-1}), but for $N=3$.}
\end{figure}

In the following section, we shall discuss the collective resonance fluorescence spectrum, given by the expression (\ref{fSp}), of a 
collection of closely packed and dipole-dipole interacting two-level emitters driven by a moderately intense and resonant coherent 
laser field.

\section{Results and Discussion \label{RD}}
\begin{figure}[t]
\includegraphics[width =7cm]{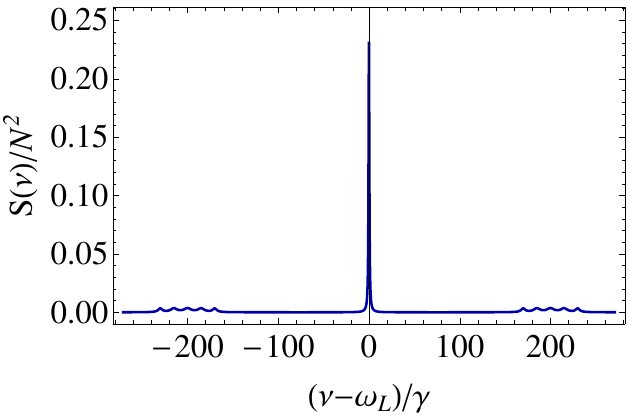}
\includegraphics[width =7cm]{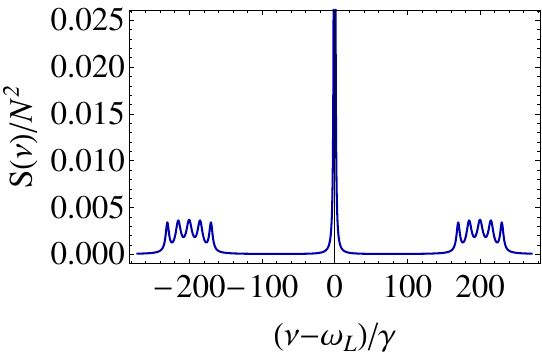}
\begin{picture}(0,0)
\put(-160,250){(a)}
\put(-150,115){(b)}
\end{picture}
\caption{\label{fig-3}
(a) The scaled cooperative resonance fluorescence spectrum $S({\nu})/N^{2}$ as a function of $(\nu -\omega_{L})/\gamma$.
Here $N=5$, $2\Omega/\gamma=200$, $\Delta/\gamma=0$, and $\delta/(2\Omega)=0.3$, respectively. (b) Same as in (a) but
for lower values on the Y-axes.}
\end{figure}
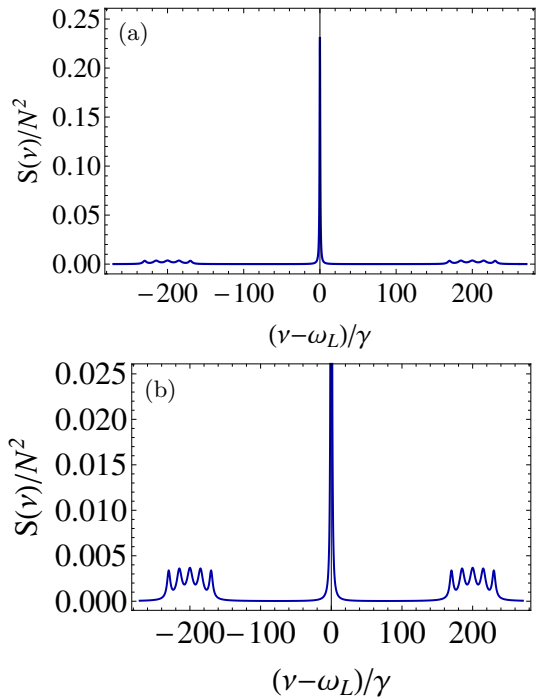

While for $\delta/2\Omega \to 0$ we have obtained that the collective resonance fluorescence spectrum consists from three-spectral 
lines, similar to the Mollow spectrum, things are different if $\delta/2\Omega \not=0$, so that $\delta/2\Omega < 1$, as it is the case 
discussed here. Actually, additional spectral sidebands appear which are centered around $\omega_{L} \pm 2\Omega$. Evidently,
this occurs due to the dipole-dipole interaction among the two-level emitters. In this context, Fig.~(\ref{fig-1}) shows the incoherent 
resonance fluorescence spectrum for a two-atom system, in the Dicke limit, i.e. the space-interval between the two atoms is much 
smaller than the corresponding photon emission wavelength, but with the dipole-dipole interaction taken into account, respectively. 
The spectrum consists from symmetrically located five spectral lines at $\nu=\omega_{L}$ and $\nu-\omega_{L} = \pm 2\Omega \pm \delta/2$, 
see also Appendix~\ref{appB} and \cite{kilin2N,helen,twat1,twat2}. The frequency interval among the two spectral-lines, around 
$\nu-\omega_{L} = \pm 2\Omega$, equals the dipole-dipole coupling strength $\delta$. An explanation of the resonance fluorescence 
spectrum given in Fig.~(\ref{fig-1}) can be done using the double-dressed state formalism, see Fig.~(\ref{fig-0ab}): The laser-emitter 
dressed states $|\tilde e\rangle_{j}$ and $|\tilde g\rangle_{j}$ additionally split due to the dipole-dipole interaction leading to the 
collective two-atom dressed-states, see e.g. \cite{mmchk}, which are responsible for the spontaneously emitted spectrum. On the 
other side, Fig.~(\ref{fig-2}) depicts the incoherent resonance fluorescence spectrum of resonantly driven $N=3$ two-level emitters, 
also for $\Delta/\gamma=0$ and within the Dicke-limit with dipole-dipole interaction being included, respectively. This time, the spectrum 
consists of seven spectral lines detected, correspondingly, at $\nu=\omega_{L}$, $\nu=\omega_{L} \pm 2\Omega$ and 
$\nu-\omega_{L}= \pm 2\Omega \pm \delta/2$, see also Appendix~\ref{appC}. Again, here, the three-spectral lines around 
$\nu-\omega_{L}=\pm 2\Omega$ are localized within the dipole-dipole coupling strength $\delta$, see Fig.~(\ref{fig-2}). In the same 
vein, a $N=5$ resonantly pumped two-level sample generates 11 spectral lines. The incoherent resonant fluorescence spectrum is 
symmetrically located with respect to the central line at $\nu=\omega_{L}$, see Fig.~(\ref{fig-3}). Each of five spectral side-bands 
are generated around $\nu-\omega_{L}=\pm2\Omega$ in a frequency range equal to the dipole-dipole coupling strength $\delta$. 

Generalizing, one can observe that the cooperative resonance fluorescence spectrum of a moderately laser driven small two-level ensemble, 
within the Dicke-limit with dipole-dipole interaction included, consists of $2N+1$ spectral lines. A central line at $\nu=\omega_{L}$, and 
$2N$ spectral lines symmetrically generated around the frequencies $\nu-\omega_{L}=\pm 2\Omega$, respectively, see also Fig.~(\ref{fig-0ab}).
The sidebands occurs due to transitions among the $N+1$ symmetrical collective Dicke states formed from individual laser-atom dressed 
states, see also the Appendix \ref{appB} and the Appendix \ref{appC}. Remarkably here, one can estimate the number of involved 
two-level emitters via detection of the incoherent collective resonance fluorescence spectrum, because both sidebands consists from 
$N$ spectral lines. The frequency separation among these $N$ spectral lines, each generated on both sides of the spectrum with 
respect to the central line at $\nu=\omega_{L}$, is equal to $\delta/(N-1)$. These lines are distinguishable if $\delta/(N-1) \gg \gamma$ 
when $\Delta/\gamma=0$ or, equivalently, $\omega_{L} \equiv \omega_{0}+\tilde \delta$. Knowing the dipole-dipole coupling strength 
from the spectrum, one can extract the mean-distance among the closely spaced two-level radiators, because the dipole-dipole interaction 
scales inversely proportional to the cubic mean inter-particle separations, respectively, see also \cite{JorgZS}. Note that Ref.~\cite{kilin2N} 
concludes too that an increase of the number of interacting atoms leads to an increase in the number of additional spectral components.

As the spectral lines are well separated for $2\Omega > \delta \gg N\gamma$, one can conjecture then, that larger atomic ensembles, with 
a fixed ratio of $2\Omega/N\gamma \gg 1$, would lead to a three spectral-line Mollow spectrum, where the two sidebands, generated at 
$\nu=\omega_{L} \pm 2\Omega$, broadens if $\delta/(N-1) \sim \gamma$, i.e. the spectral sidebands overlap in this case. The frequency 
bandwidth of these two spectral lines, located at $\nu=\omega_{L} \pm 2\Omega$ in the Mollow like-spectrum, would be close to the 
dipole-dipole coupling strength, $\delta$, see also Fig.~\ref{fig-3}(b) where one can anticipate that the sidebands would overlap for lower 
values of the dipole-dipole coupling strengths. Notice that sidebands broadening of the Mollow spectrum, in a regular sub-wavelength chain 
\cite{colar} of laser pumped dipole-dipole interacting two-level atoms, was recently reported as well in Ref.~\cite{DDLchain}.

\section{Summary \label{sum}}
We have investigated the collective steady-state quantum dynamics of an externally resonantly pumped two-level ensemble, 
concentrated in a small volume within the Dicke limit and secular approximation. However, the dipole-dipole interactions among 
the emitters are taken into account, while their coupling strength is considered to be of the order of the corresponding Rabi 
frequency, but still smaller. As a result, we have found that the incoherent collective resonance fluorescence spectrum consists 
of multiple spectral lines which are dependent on ensemble's sizes. As a consequence, one can estimate the number of involved 
emitters via measuring the spontaneously scattered resonance fluorescence spectrum. This is feasible since the number of the 
spectral sidebands, arising due to transitions among the N+1 symmetrized collective Dicke dressed states, equals to the doubled 
number of two-level radiators within the laser-pumped sample. Actually, these spectral sidebands are distinguishable if the 
dipole-dipole coupling is larger than the collective spontaneous decay rate. In the opposite case, the incoherent collective 
resonance fluorescence spectrum is formed from three lines, similar to the Mollow spectrum, whereas the sidebands spectral 
lines broadens proportional to the dipole-dipole coupling strength, respectively. In both case, one can estimate the dipole-dipole 
coupling strength as well as the mean-distance among the closely spaced laser-driven two-level emitters.

\acknowledgments
The financial support from the Moldavian Ministry of Education and Research, via grant No. 011205, is gratefully acknowledged.

\appendix

\section{Determining the decay rate of the off-diagonal matrix elements \label{appA}}
The equations of motion entering in the expression for the collective resonance fluorescence spectrum (\ref{sp1}), can be
easily obtained from the master equation (\ref{fMeq}) for $\theta=\pi/4$ and $\delta=0$ only. Otherwise, those equations
of motion will involve higher order atomic correlators leading to difficulties in solving them, even numerically, for $N \gg 1$.
In the following, we shall determine the decay rates of the off-diagonal terms of the master equation (\ref{fMeq}), for 
$\theta=\pi/4$ and $\delta/2\Omega<1$. In the secular approximation, that is $2\Omega > \delta \gg N\gamma$, there 
are $N+1$ distinguishable collective Dicke ladder laser-atom dressed states, see also Appendixes \ref{appB} and \ref{appC} 
where this is demonstrated, particularly, for $N=2$ and $N=3$ atomic samples. Then, the time-behaviours of the off-diagonal 
elements, i.e. $\langle n|\rho|n+1\rangle$, of the maser equation (\ref{fMeq}),
\begin{eqnarray}
\frac{d}{dt}\rho(t) &+& i[\bar G R_{z}-\bar\delta R^{+}R^{-},\rho]= \nonumber \\ 
&-&\frac{\gamma}{8}\bigl([R_{z},R_{z}\rho] + [R^{+},R^{-}\rho] + [R^{-},R^{+}\rho] \bigr), \nonumber \\
&+& H.c. \label{meqr}
\end{eqnarray}
are given, respectively, as follows
\begin{eqnarray}
&{}&\frac{d}{dt}\rho_{n,n+1}= -i\biggl(2\Omega-\frac{\tilde\delta}{2}\bigl(1+(2n-N)\bigr)\biggr)\rho_{n,n+1} \nonumber \\
&-&\frac{\gamma}{4}\biggl(1+ 2(N-n)(n+1)\biggr)\rho_{n,n+1} \nonumber \\
&+& \frac{\gamma}{4}\biggl(\sqrt{n(n+1)(N-n+1)(N-n)}\rho_{n-1,n} \nonumber \\
&+&\sqrt{(n+1)(n+2)(N-n-1)(N-n)}\rho_{n+1,n+2}\biggr), \nonumber \\ 
\label{offd}
\end{eqnarray}
where $|n\rangle$ are the symmetrical collective atomic states, and we used that $\bar G=\Omega-\tilde\delta/4$ as well 
as $\bar\delta=-\tilde\delta/2$ when $\theta=\pi/4$. One observes here that the decay rates of the off-diagonal elements 
are given by the second line of Eqs.~(\ref{offd}), i.e. $\gamma_{n} \equiv \gamma(1+2(N-n)(n+1))/4$, or in the operator
form as: $\hat \gamma_{c}=\gamma(1+2R^{-}R^{+})/4$. When combining with the coherent part of Eqs.~(\ref{offd}),
we obtain: $R^{-}(\tau)=e^{-(i\hat\omega + \hat\gamma_{c})\tau}R^{-}$, $R^{+}(\tau)=R^{+}e^{(i\hat\omega - 
\hat\gamma_{c})\tau}$, with $\hat \omega=2\Omega - \tilde\delta(1+R_{z})/2$, which are given in the Exps.~(\ref{td}).

\section{The resonance fluorescence spectrum for a dipole-dipole interacting pair of two-level emitters \label{appB}}
Here we shall obtain the cooperative resonance fluorescence spectrum of a resonantly laser-pumped pair of dipole-dipole coupled 
two-level radiators, directly from the master equation (\ref{fMeq}), when $\theta=\pi/4$. Introducing the collective two-atom 
dressed states \cite{lehm}: 
$|E\rangle = |\tilde e_{1}\tilde e_{2}\rangle$, $|S\rangle=\bigl(|\tilde e_{1}\tilde g_{2}\rangle + |\tilde e_{2}\tilde g_{1}\rangle\bigr)/\sqrt{2}$,
$|A\rangle=\bigl(|\tilde e_{1}\tilde g_{2}\rangle - |\tilde e_{2}\tilde g_{1}\rangle\bigr)/\sqrt{2}$, and $|G\rangle = |\tilde g_{1}\tilde g_{2}\rangle$,
and taking into account that $R^{+}_{1}=\bigl(\bar R_{ES} - \bar R_{EA} + \bar R_{SG} + \bar R_{AG}\bigr)/\sqrt{2}$ and 
$R^{+}_{2}=\bigl(\bar R_{ES} + \bar R_{EA} + \bar R_{SG} - \bar R_{AG}\bigr)/\sqrt{2}$ one obtains from (\ref{fMeq}) if $\theta=\pi/4$, the 
following master equation in the cooperative two-atom dressed states bases,
\begin{eqnarray}
\frac{d}{dt}\rho(t) &+& \frac{i}{\hbar}[\bar H,\rho]= -\frac{\gamma}{2}[(\bar R_{EE} - \bar R_{GG}),(\bar R_{EE} - \bar R_{GG})\rho] 
\nonumber \\
&-&\frac{\gamma}{4}\biggl([\bar R_{ES},\bar R_{SE}\rho] + [\bar R_{SG},\bar R_{GS}\rho] + [\bar R_{SE},\bar R_{ES}\rho] 
\nonumber \\
&+& [\bar R_{GS},\bar R_{SG}\rho] \biggr) + H.c..
\label{MeqA}
\end{eqnarray}
Here $\bar H=2\hbar\Omega(\bar R_{EE} - \bar R_{GG})+\hbar\tilde \delta \bar R_{SS}/2$, while $\bar R_{\alpha\beta}=|\alpha\rangle \langle \beta|$ 
and $[\bar R_{\alpha\beta},\bar R_{\beta'\alpha'}]=\bar R_{\alpha\alpha'}\delta_{\beta\beta'} - \bar R_{\beta'\beta}\delta_{\alpha'\alpha}$ with 
$\{\alpha, \beta \in E, S, G \}$. The anti-symmetrical state $|A\rangle$ was dropped since does not participate in the dynamics within the Dicke limit.
Furthermore we have assumed that $2\Omega > \delta \gg N\gamma$ meaning that the corrections to the resonance fluorescence spectrum are 
of the order of $(\delta/2\Omega)^{2}$. Respectively, the resonance fluorescence spectrum (\ref{sp1}) takes the next form in the collective two-atom 
dressed states basis:
\begin{eqnarray}
S(\nu)&=&\frac{1}{4}{\rm Re}\int^{\infty}_{0}d\tau e^{i(\nu-\omega_{L})\tau} \nonumber \\
&\times& \biggl\{4\langle \bigl(\bar R_{EE}-\bar R_{GG}\bigr)\bigl(\bar R_{EE}(\tau)-\bar R_{GG}(\tau)\bigr)\rangle_{s} 
\nonumber \\
&+& 2\bigl( \langle \bar R_{ES}\bar R_{SE}(\tau)\rangle_{s} + \langle \bar R_{SG}\bar R_{GS}(\tau)\rangle_{s} 
\nonumber \\
&+& \langle \bar R_{SE}\bar R_{ES}(\tau)\rangle_{s} + \langle \bar R_{GS}\bar R_{SG}(\tau)\rangle_{s}\bigr) \biggr \}.
\label{spto}
\end{eqnarray}
Now, from the master equation (\ref{MeqA}), it is easily to obtain and solve the equations of motion entering in the above expression (\ref{spto}), 
that is,
\begin{eqnarray}
\bar R_{EE}(\tau) - \bar R_{GG}(\tau) = \bigl(\bar R_{EE}(0)-\bar R_{GG}(0)\bigr)e^{-\gamma\tau/2}, \nonumber \\
\bar R_{SE}(\tau)=\bar R_{SE}(0)e^{-\{i(2\Omega-\tilde\delta/2)+5\gamma/4\}\tau},\nonumber \\
\bar R_{GS}(\tau)=\bar R_{GS}(0)e^{-\{i(2\Omega+\tilde\delta/2)+5\gamma/4\}\tau}. \nonumber \\
\label{eqmA}
\end{eqnarray}
After substitution of time-dependent solutions (\ref{eqmA}) in the expression (\ref{spto}) and using the quantum regression theorem, one obtains the 
following expression for the two-atom resonance fluorescence spectrum:
\begin{eqnarray}
S(\nu)&=&\frac{1}{6}\biggl\{4\frac{\gamma/2}{(\gamma/2)^{2} + (\nu-\omega_{L})^{2}} \nonumber\\
&+& \frac{5\gamma/4}{(5\gamma/4)^{2} 
+ (\nu-\omega_{L}-2\Omega - \delta/2)^{2}} \nonumber \\
&+& \frac{5\gamma/4}{(5\gamma/4)^{2} + (\nu-\omega_{L}-2\Omega+ \delta/2)^{2}} \nonumber \\
&+& \frac{5\gamma/4}{(5\gamma/4)^{2} + (\nu-\omega_{L}+2\Omega+ \delta/2)^{2}} \nonumber\\
&+&\frac{5\gamma/4}{(5\gamma/4)^{2} + (\nu-\omega_{L}+2\Omega- \delta/2)^{2}}\biggr\}, \label{toSp} 
\end{eqnarray}
where we have used that: $\langle \bar R_{EE}\rangle_{s}= \langle \bar R_{SS}\rangle_{s} = \langle \bar R_{GG}\rangle_{s}=1/3$.
One observes that the resonance fluorescence spectrum for a two-atom system, within the Dicke limit, consists of five spectral lines, 
see Fig.~\ref{fig-0ab}(b) and Fig.~(\ref{fig-1}). One central line at $\nu=\omega_{L}$ and four spectral lines at $\nu-\omega_{L} 
= \pm 2\Omega \pm \tilde\delta/2$, respectively. The sideband spectral lines are due to transitions among the two-atom cooperative 
dressed-states, i.e., $|E\rangle \leftrightarrow |S\rangle \leftrightarrow |G\rangle$. As there are involved only the symmetrical two-atom 
Dicke states with two allowed transitions among them in both directions, one has $2\times 2=4$ sidebands, or generally $2N$. This will 
be also the case for $N=3$, see Appendix \ref{appC}. The central line appears because of the transitions among the two-atom dressed-states 
$|E\rangle \leftrightarrow |E\rangle$ and $|G\rangle \leftrightarrow |G\rangle$, see Fig.~\ref{fig-0ab}(b), and, hence, there are in total 
$2\times2 +1=5$ spectral lines for a two-atom sample. Note that, because we have assumed that $\delta \gg \gamma$ one can not 
recover, from (\ref{toSp}), the two-atom spectrum given by Exp.~(\ref{fSd0}) for $\delta=0$ when setting $N=2$. Finally, we 
emphasise that Fig.~(\ref{fig-0ab}) is just a scheme that intuitively may help to understand the N-atom resonance fluorescence 
spectrum and should be treated correspondingly.

\section{The resonance fluorescence spectrum for $N=3$ dipole-dipole interacting two-level emitters \label{appC}}
We proceed by giving the three-atom collective dressed-states,  see e.g. \cite{thrF}: $|8\rangle=|\tilde e_{1}\tilde e_{2}\tilde e_{3}\rangle$,
$|7\rangle=\bigl(|\tilde g_{1}\tilde e_{2}\tilde e_{3}\rangle - |\tilde e_{1}\tilde e_{2}\tilde g_{3}\rangle\bigr)/\sqrt{2}$,
$|6\rangle=\bigl(-|\tilde g_{1}\tilde e_{2}\tilde e_{3}\rangle + 2|\tilde e_{1}\tilde g_{2}\tilde e_{3}\rangle - 
|\tilde e_{1}\tilde e_{2}\tilde g_{3}\rangle\bigr)/\sqrt{6}$, $|5\rangle=\bigl(|\tilde g_{1}\tilde e_{2}\tilde e_{3}\rangle 
+ |\tilde e_{1}\tilde g_{2}\tilde e_{3}\rangle + |\tilde e_{1}\tilde e_{2}\tilde g_{3}\rangle\bigr)/\sqrt{3}$,
$|4\rangle=\bigl(|\tilde e_{1}\tilde g_{2}\tilde g_{3}\rangle - |\tilde g_{1}\tilde g_{2}\tilde e_{3}\rangle\bigr)/\sqrt{2}$,
$|3\rangle=\bigl(-|\tilde e_{1}\tilde g_{2}\tilde g_{3}\rangle + 2|\tilde g_{1}\tilde e_{2}\tilde g_{3}\rangle - 
|\tilde g_{1}\tilde g_{2}\tilde e_{3}\rangle\bigr)/\sqrt{6}$,
$|2\rangle=\bigl(|\tilde e_{1}\tilde g_{2}\tilde g_{3}\rangle + |\tilde g_{1}\tilde e_{2}\tilde g_{3}\rangle + 
|\tilde g_{1}\tilde g_{2}\tilde e_{3}\rangle\bigr)/\sqrt{3}$, and $|1\rangle=|\tilde g_{1}\tilde g_{2}\tilde g_{3}\rangle$. 
Then, the master equation (\ref{fMeq}) takes the following form in the three-atom dressed-state bases, namely,
\begin{eqnarray}
\frac{d}{dt}\rho(t)&+&\frac{i}{\hbar}[\tilde H,\rho]= -\frac{\gamma}{2}[\tilde R_{0},\tilde R_{0}\rho] -\frac{3\gamma}{8}\biggl(
[\tilde R_{21},\tilde R_{12}\rho] \nonumber \\
&+& [\tilde R_{85},\tilde R_{58}\rho] + [\tilde R_{12},\tilde R_{21}\rho] + [\tilde R_{58},\tilde R_{85}\rho]\biggr) \nonumber \\
&-& \frac{\gamma}{8}\biggl([\tilde R^{+},\tilde R^{-}\rho] + [\tilde R^{-},\tilde R^{+}\rho] \biggr), \label{MeqT}
\end{eqnarray}
where we have used that: $R^{+}_{1}=(\tilde R_{41}+\tilde R_{87})/\sqrt{2}+(\tilde R_{21}-\tilde R_{64} - \tilde R_{73} + 
\tilde R_{85})/\sqrt{3} - (\tilde R_{31}+\tilde R_{54} + \tilde R_{72} + \tilde R_{86})/\sqrt{6} - (\tilde R_{53}+\tilde R_{62})/\sqrt{18}
+ \sqrt{2}(\tilde R_{62} + \tilde R_{53})/3 + 2(\tilde R_{52}-\tilde R_{63})/3$, while $R^{+}_{2}=\sqrt{2/3}(\tilde R_{31}+\tilde R_{86})
+ (\tilde R_{21}+\tilde R_{85})/\sqrt{3}-\sqrt{2}(\tilde R_{62}+\tilde R_{53})/3+2(\tilde R_{63}/2 + \tilde R_{52})/3 - \tilde R_{74}$ and
$R^{+}_{3}=-(\tilde R_{41}+\tilde R_{87})/\sqrt{2}+(\tilde R_{21}+\tilde R_{64}+\tilde R_{73} + 
\tilde R_{85})/\sqrt{3} + (\tilde R_{54} + \tilde R_{72} - \tilde R_{31}-\tilde R_{86})/\sqrt{6} - (\tilde R_{53}+\tilde R_{62})/\sqrt{18}
+ \sqrt{2}(\tilde R_{62} + \tilde R_{53})/3 + 2(\tilde R_{52}-\tilde R_{63})/3$. Also, $R^{-}_{j}=[R^{+}_{j}]$, $\{j \in 1,2,3\}$. 

In the master equation (\ref{MeqT}), $\tilde H/\hbar=(2\Omega+\tilde\delta)\tilde R_{22}+(2\Omega-\tilde\delta/2)\tilde R_{33}
+(2\Omega-\tilde\delta/2)\tilde R_{44}+(4\Omega+\tilde\delta)\tilde R_{55}+(4\Omega-\tilde\delta/2)\tilde R_{66} + 
(4\Omega-\tilde\delta/2)\tilde R_{77}+6\Omega\tilde R_{88}$, whereas $\tilde R_{0}=\tilde R_{22} + \tilde R_{33} + \tilde R_{44}+
2(\tilde R_{55}+\tilde R_{66}+\tilde R_{77})+3\tilde R_{88}$ and $\tilde R^{+}=2\tilde R_{52}-\tilde R_{63}-\tilde R_{74}$, with 
$\tilde R^{-}=[\tilde R^{+}]^{\dagger}$. The three-atom dressed-states operators are defined as follows: 
$\tilde R_{\alpha\beta}=|\alpha\rangle\langle \beta|$ and satisfying the commutation relations $[\tilde R_{\alpha\beta},\tilde R_{\beta'\alpha'}]
=\tilde R_{\alpha\alpha'}\delta_{\beta\beta'} - \tilde R_{\beta'\beta}\delta_{\alpha'\alpha}$ where
$\{\alpha, \beta \in 1, \cdots, 8\}$. The master equation (\ref{MeqT}) involves symmetrical as well as anti-symmetrical three-atom collective 
dressed-states, respectively. Fortunately, only the symmetrical states will contribute to the three-atom resonance fluorescence spectrum. 
In this regard, the resonance fluorescence spectrum, represented via the three-atom cooperative dressed-states, is given by
\begin{eqnarray}
S(\nu)&=&\frac{1}{4}{\rm Re}\int^{\infty}_{0}d\tau e^{i(\nu-\omega_{L})\tau} \nonumber \\
&\times& \biggl\{\langle (2\tilde R_{0}-3)(2\tilde R_{0}(\tau)-3)\rangle_{s}  + 3\bigl(\langle\tilde R_{21}\tilde R_{12}(\tau)\rangle_{s} \nonumber \\
&+& \langle\tilde R_{85}\tilde R_{58}(\tau)\rangle_{s} + \langle\tilde R_{12}\tilde R_{21}(\tau)\rangle_{s} 
+ \langle\tilde R_{58}\tilde R_{85}(\tau)\rangle_{s} \bigr) \nonumber \\
&+& \langle\tilde R^{+}\tilde R^{-}(\tau)\rangle_{s} + \langle\tilde R^{-}\tilde R^{+}(\tau)\rangle_{s} \biggr\}. \label{spthr}
\end{eqnarray}
The equations of motion necessary to calculate the spectrum can be obtained directly from the master equation (\ref{MeqT}). In the following 
we give their solutions, that is,
\begin{eqnarray}
\tilde R_{0}(\tau)&=&\tilde R_{0}(0)e^{-\gamma\tau/2}+3(1-e^{-\gamma\tau/2})/2, \nonumber \\
\tilde R_{12}(\tau)&=&\tilde R_{12}(0)e^{-\{i(2\Omega+\tilde\delta) + 7\gamma/4\}\tau}, \nonumber \\
\tilde R_{58}(\tau)&=&\tilde R_{58}(0)e^{-\{i(2\Omega-\tilde\delta) + 7\gamma/4\}\tau}, \nonumber \\
\tilde R_{25}(\tau)&=&\tilde R_{25}(0)e^{-\{2i\Omega + 9\gamma/4\}\tau}, \nonumber \\
\tilde R_{36}(\tau)&=&\tilde R_{36}(0)e^{-\{2i\Omega +3\gamma/4\}\tau}, \nonumber \\
\tilde R_{47}(\tau)&=&\tilde R_{47}(0)e^{-\{2i\Omega +3\gamma/4\}\tau}. \label{sthr}
\end{eqnarray}
Taking into account that when all the three atoms are initially in their bare ground states, then one has: $\langle\tilde R_{33}(0)\rangle
=\langle\tilde R_{44}(0)\rangle=\langle\tilde R_{66}(0)\rangle=\langle\tilde R_{77}(0)\rangle =0$ and so does their steady-state expectation 
values. This way, the anti-symmetrical three-atom dressed-states, i.e. $|\alpha\rangle$ with $\{\alpha \in 3,4,6,7\}$, do not contribute to the 
final expression of the resonance fluorescence spectrum. After substitution of solutions (\ref{sthr}) in Exp.~(\ref{spthr}), and using the 
quantum regression theorem, one arrives at the following expression for the laser-pumped three-atom resonance fluorescence spectrum, 
namely,
\begin{widetext}
\begin{eqnarray}
S(\nu)&=&\frac{1}{4}\biggl\{5\frac{\gamma/2}{(\gamma/2)^{2} + (\nu-\omega_{L})^{2}} + 
\frac{9\gamma/4}{(9\gamma/4)^{2} + (\nu-\omega_{L}-2\Omega)^{2}} + \frac{9\gamma/4}{(9\gamma/4)^{2} 
+ (\nu-\omega_{L}+2\Omega)^{2}} \nonumber \\
&+& \frac{3}{4}\frac{7\gamma/4}{(7\gamma/4)^{2} + (\nu-\omega_{L}-2\Omega - \delta/2)^{2}} +
\frac{3}{4}\frac{7\gamma/4}{(7\gamma/4)^{2} + (\nu-\omega_{L}-2\Omega + \delta/2)^{2}} \nonumber \\
&+& \frac{3}{4}\frac{7\gamma/4}{(7\gamma/4)^{2} + (\nu-\omega_{L}+2\Omega + \delta/2)^{2}} +
\frac{3}{4}\frac{7\gamma/4}{(7\gamma/4)^{2} + (\nu-\omega_{L}+2\Omega - \delta/2)^{2}}\biggr\}, \label{thrSp} 
\end{eqnarray}
\end{widetext}
where we used that: $\langle\tilde R_{11}\rangle_{s}=\langle\tilde R_{22}\rangle_{s}=\langle\tilde R_{55}\rangle_{s}=
\langle\tilde R_{88}\rangle_{s} =1/4$. Here again, the sidebands arise due to transitions, in both directions, among the symmetrical three-atom 
dressed-states, i.e., $|1\rangle \leftrightarrow |2\rangle \leftrightarrow |5\rangle \leftrightarrow |8\rangle$, respectively. So, there are 
$2\times 3=6$, that is $2N$, sidebands. Generalizing, the spectral lines emitted at $\nu-\omega_{L}=\pm 2\Omega$, $\nu-\omega_{L}
=\pm (2\Omega+\tilde \delta)$, and $\nu-\omega_{L}=\pm (2\Omega-\tilde \delta)$ are due to transitions among the symmetrical collective 
states $|2\rangle \leftrightarrow |5\rangle$, $|1\rangle \leftrightarrow |2\rangle$ and $|5\rangle \leftrightarrow |8\rangle$, respectively. The 
central one occurs due to light scattering on $|\alpha\rangle \leftrightarrow |\alpha\rangle$, $\{\alpha \in 1,2,5,8\}$, three-atom collective 
dressed-states, see also Fig.~(\ref{fig-0ab}). Thus, concluding, one has a total of $2N+1$ spectral lines in resonance fluorescence processes 
involving a collection of laser-pumped $N$ dipole-dipole interacting two-level emitters, within the Dicke limit and the secular approximation, 
i.e. $2\Omega > \delta \gg N\gamma$, respectively. The sidebands occurs due to transitions among the $N+1$ symmetrical collective Dicke 
states formed from individual laser-atom dressed states.

\end{document}